\documentclass[5p]{elsarticle}
\usepackage{amsmath, calrsfs}
\usepackage{graphicx,float,color}
\usepackage{hyperref}
\hypersetup{colorlinks = true, linkcolor = blue, urlcolor  = blue, citecolor = blue, anchorcolor = blue}
\usepackage{multirow}
\usepackage{natbib}

\begin{document}

\title{The compatibility of LHC Run 1 data with a heavy scalar of mass around 270\,GeV}

\author[wits]{Stefan von Buddenbrock}
\ead{stef.von.b@cern.ch}

\author[hri]{Nabarun Chakrabarty} 
\ead{nabarunc@hri.res.in}

\author[mitp]{Alan S. Cornell}
\ead{Alan.Cornell@wits.ac.za}

\author[wits]{Deepak Kar}
\ead{Deepak.Kar@cern.ch}

\author[mitp]{Mukesh Kumar}
\ead{mukesh.kumar@cern.ch}

\author[hri]{Tanumoy Mandal} 
\ead{tanumoymandal@hri.res.in}

\author[wits]{Bruce Mellado} 
\ead{Bruce.Mellado@wits.ac.za}

\author[hri]{Biswarup Mukhopadhyaya} 
\ead{biswarup@hri.res.in}

\author[wits]{Robert G. Reed}
\ead{Robert.Reed@cern.ch}

\address[wits]{School of Physics, University of the Witwatersrand, Johannesburg, Wits 2050, South Africa}

\address[hri]{Regional Centre for Accelerator-based Particle Physics, Harish-Chandra Research Institute, Chhatnag Road, Jhusi, Allahabad - 211 019, India.}

\address[mitp]{National Institute for Theoretical Physics; School of Physics and Mandelstam Institute for Theoretical Physics, University of the Witwatersrand, Johannesburg, Wits 2050, South Africa}

\date{\today}

\begin{abstract}
The first run of the LHC was successful in that it saw the discovery of the elusive Higgs boson, a particle that is consistent with the SM hypothesis. There are a number of excesses in Run 1 ATLAS and CMS results which can be interpreted as being due to the existence of another heavier scalar particle. This particle has decay modes which we have studied using LHC Run 1 data. Using a minimalistic model, we can predict the kinematics of these final states and compare the prediction against data directly. A statistical combination of these results shows that a best fit point is found for a heavy scalar having a mass of 272$^{+12}_{-9}$\,GeV. This result has been quantified as a three sigma effect, based on analyses which are not necessarily optimized for the search of a heavy scalar. The smoking guns for the discovery of this new heavy scalar and the prospects for Run 2 are discussed.
\end{abstract}

\begin{keyword}
Higgs boson \sep Heavy scalar \sep Dark matter
\PACS 14.80.Bn \sep 14.80.Ec \sep 12.60.Cn \sep 12.60.Fr
\end{keyword}

\maketitle

\section{Introduction\label{sec:intro}}

With the discovery of a new scalar boson (which will be denoted
by $h$) at the Large Hadron Collider
(LHC)~\cite{Aad:2012tfa,Chatrchyan:2012xdj}, new tasks and explorations
have come to the fore. Both the ATLAS and CMS experiments have been keenly exploring the properties of the new scalar, and will continue
to do so in the high-energy run(s). This includes tests on various coupling strengths as well as
spin-CP properties. Overall, the global fits of the data indicate that the properties are mostly
consistent with what the Standard Model (SM) of particle physics expects for a Higgs boson. 
Nonetheless, there are certain features and excesses in the available data that warrant detailed attention. Table~\ref{tab:list_of_results} summarizes the measurements made by the ATLAS and CMS experiments that are considered here. These results comprise the measurements of the differential Higgs boson transverse momentum ($p_{Th}$) (where $h$ is the SM-like Higgs.), searches for a di-Higgs boson resonance, the Higgs boson in association with top quarks, and $VV$ resonances (where $V=Z,W^\pm$). These final states are considered against the hypothesis of a heavy scalar boson ($H$). 

In order to describe the shape of the $p_{Th}$ distribution, it is necessary to introduce decays in which at least one $h$ is produced: $H\rightarrow hh, h\chi\chi$ where $\chi$ would be a dark matter candidate, leading to the production of the Higgs boson in association with missing energy. The latter could be realized through the decay of some intermediate particle. This hypothetical intermediate particle (the existence or nature of which we make no statement about) could also decay into a pair of light hadronic jets. In either case, a distortion of the $p_{Th}$ spectrum would be expected.

\begin{table*}
	\renewcommand{\arraystretch}{1.25}
	\centering
	\begin{tabular}{|c|c|l|}
		\hline
		Result & \multicolumn{2}{c|}{Publication} \\
		\hline \hline
		\multirow{2}{100pt}{\centering Differential Higgs boson $p_T$ spectra} & \textbf{ATLAS} & Fiducial cross section measurements on $h\to\gamma\gamma$~\cite{Aad:2014lwa} and $h\to ZZ^*\to4\ell$~\cite{Aad:2014tca} \\
		\cline{2-3}
		& \textbf{CMS} & Fiducial cross section measurements on $h\to\gamma\gamma$~\cite{Khachatryan:2015rxa} and $h\to ZZ^*\to4\ell$~\cite{CMS:2015hja}  \\
		\hline
		\multirow{2}{100pt}{\centering Di-Higgs boson resonance searches}  & \multirow{1}{*}{\textbf{ATLAS}} & Limits on $H\to hh\to b\bar{b}\tau\tau$, $\gamma\gamma WW^*$, $\gamma\gamma b\bar{b}$ and $b\bar{b}b\bar{b}$~\cite{Aad:2015xja} \\
		\cline{2-3}
		& \textbf{CMS} & Limits on $H\to hh\to\gamma\gamma b\bar{b}$~\cite{CMS:2014ipa}, $b\bar{b}\tau\tau$~\cite{Khachatryan:2015tha} and multi-lepton~\cite{Khachatryan:2014jya} \\
		\hline
		\multirow{3}{100pt}{\centering Top associated Higgs boson production} & \multirow{2}{*}{\textbf{ATLAS}} & Limits on $h\to\gamma\gamma$~\cite{Aad:2014lma} \\
		& & Measurements on multi-lepton decay channels~\cite{Aad:2015iha} and $h\to b\bar{b}$~\cite{Aad:2015gra}  \\
		\cline{2-3}
		& \textbf{CMS} & Measurements on $h\to\gamma\gamma$, $h\to b\bar{b}$ and multi-lepton decay channels~\cite{Khachatryan:2014qaa} \\
		\hline
		\multirow{2}{100pt}{\centering $H\to VV$ decays} & \multirow{1}{*}{\textbf{ATLAS}} & Limits on $H\to WW$~\cite{Aad:2015agg} and $ZZ$~\cite{Aad:2015kna} \\
		\cline{2-3}
		& \textbf{CMS} & Limits on $H\to WW$ and $ZZ$~\cite{Khachatryan:2015cwa} \\
		\hline
		
	\end{tabular}
	\caption{A list of the experimental results which were used to help constrain the relevant parameters of the proposed model. In the interest of being as unbiased as possible, these results were selected regardless of whether they hint at physics beyond the SM.}
	\label{tab:list_of_results}
\end{table*}

Both ATLAS and CMS have observed excesses in the production of a SM Higgs boson in association with top quarks. The new boson would naturally be produced in association with top quarks. In addition, with a small branching ratio to $VV$, the effect of negative interference in single top associated production is suppressed, and we would expect a large cross section for a heavy scalar being produced in association with top quarks. These effects would yield an explanation for the excesses which are seen in the data.

While none of the excesses summarized in Table~\ref{tab:list_of_results} is significant enough on its own, it is tantalizing to study their compatibility as a whole with the decay of one new heavy boson. In fact, viewed under this hypothesis, the overall size of the excess could be significant. We adopt a bottom-up approach and wait for an overseeing theory until
the observed results are further consolidated. The fit of the available data presented here by us will hopefully enable people to put all pieces of the jigsaw puzzle together once a larger volume of results accumulates. It is important to note that some of the experimental analyses used in this study were not optimized for the search of a heavy scalar boson. Therefore, the sensitivity to the search is not maximized, rendering conservative the results reported here. 

The salient features of the suggested scenario are summarised in Section~\ref{sec:scenario}. In Sections~\ref{sec:stats} and \ref{sec:tools} the main tools of our analysis are discussed with Section~\ref{sec:stats} focussed on the statistical machinery used. The results are presented and discussed in Section~\ref{sec:results} including a discussion on smoking guns and prospects for Run 2, followed by a brief conclusion in Section~\ref{sec:conclusions}.

\section{A suggested scenario\label{sec:scenario}}
The features and excesses which are seen in the data have been treated in such a way that they could be explained purely by physics beyond the SM (BSM).
We propose a scenario which is a BSM extension to the SM, allowing us to write
\begin{equation}
\mathcal{L} = \mathcal{L}_{\text{SM}} + \mathcal{L}_{\text{BSM}},
\end{equation}
where all of the new interactions and states are encoded in $\mathcal{L}_{\text{BSM}}$.

The simplest approach is to treat the new interactions as arising from effective couplings. We have assumed $h$ to have SM interactions with fermions and gauge bosons. The sectors of the proposed BSM Lagrangian involving the new scalars (omitting
the usual mass and kinetic energy terms) include,
\begin{equation}
\mathcal{L}_{\text{BSM}} \supset \mathcal{L}_{H} + \mathcal{L}_{\text{Y}} + \mathcal{L}_{\text{T}} + \mathcal{L}_{\text{Q}},
\end{equation}
where terms $\mathcal{L}_{\text{Y}}$,
$\mathcal{L}_{\text{T}}$ and $\mathcal{L}_{\text{Q}}$ are the Yukawa, trilinear and
quartic interactions relevant for this study, respectively. These sectors are defined as follows,
\allowdisplaybreaks[4]
\begin{align}
\mathcal{L}_{H} &= -\frac{1}{4}~\beta_{g} \kappa_{_{hgg}}^{\text{SM}}~G_{\mu\nu}G^{\mu\nu}H
+\beta_{_V}\kappa_{_{hVV}}^{\text{SM}}~V_{\mu}V^{\mu}H, \label{eq:lagH}\\
\mathcal{L}_{\text{Y}} &= -\frac{1}{\sqrt{2}}~\Big[y_{_{ttH}}\bar{t} t H + y_{_{bbH}} \bar{b} b H\Big],\\ 
\mathcal{L}_{\text{T}} &=-\frac{1}{2}~v\Big[\lambda_{_{Hhh}}Hhh + \lambda_{_{h\chi\chi}}h\chi\chi + \lambda_{_{H\chi\chi}}H\chi\chi\Big], \\
\mathcal{L}_{\text{Q}} &= -\frac{1}{2}\lambda_{_{Hh\chi\chi}}Hh\chi\chi - \frac{1}{4} \lambda_{_{HHhh}}HHhh \nonumber \\
&\hspace{0.35cm}-\frac{1}{4}\lambda_{_{hh\chi\chi}}hh \chi\chi -\frac{1}{4} \lambda_{_{HH\chi\chi}}HH\chi\chi,
\end{align}
where $H$ and $\chi$ denote the heavy scalar and the DM candidate respectively (the latter is assumed to be a scalar for illustration), and $v=246$\,GeV is the vacuum expectation value that is responsible for the $W$- and $Z$-boson masses. This can be looked upon as a variant of Higgs boson portal scenarios~\cite{Guo:2010hq,Cai:2011kb,Gonderinger:2012rd}.
This Lagrangian could in principle emerge as an effective theory after electroweak symmetry breaking in any gauge-invariant extended scalar sector. The second term in Eq.~(\ref{eq:lagH}) is summed over the weak vector bosons $Z$ and $W^\pm$, and the $\kappa$ factors are
the SM-like couplings, with $\kappa^{\text{SM}}_{_{hgg}} = \alpha_s/(3\pi v)$ and $\kappa^{\text{SM}}_{_{hVV}} \simeq m_V^2/v$.

We deliberately make no statement about the
gauge quantum numbers carried by $H$, but just postulate that the
above terms stay after electroweak symmetry breaking. We set $\beta_g \kappa^{\text{SM}}_{_{hgg}}$ to be the strength of the effective gluon-gluon coupling of $H$. In situations where there are no additional effects over and above the top-mediated
triangle diagrams contributing to this effective interaction, $\beta_g = y_{_{ttH}}/y_{_{tth}}$ where $y_{_{tth}}$ is the SM top Yukawa coupling. There would also be a similar relation for the bottom Yukawa coupling $y_{_{bbH}}$, but this has been counted as negligible since the effect of bottom quarks in gluon fusion loops is small.
The production
of $H$ is made to occur through gluon fusion and its rate can therefore be
controlled by varying $\beta_{g}$. Likewise, the $HVV$ couplings can be tuned
by varying $\beta_{_V}$.

It should again be remembered that we are not making any definite statement on the origin of each term. Thus, as we shall see below, the $H h \chi \chi$ (effective) coupling is required to be on the high side, keeping the observed data in mind. We do not rule out the possibility of this being due to the participation of some real particle in the intermediate state, as discussed above.

Major constraints on the Lagrangian parameters stem from the
observations of the relic density of DM~\cite{Ade:2013zuv} and the
DM-nuclei inelastic scattering cross sections~\cite{Akerib:2013tjd}. These constraints are controlled by the two model
parameters $m_{\chi}$ and $\lambda_{h\chi\chi}$. It is found that
both of these DM constraints can simultaneously be satisfied for a narrow
choice of the parameters $m_\chi \sim [55-60]$~GeV for very small
$\lambda_{h\chi\chi}\sim [0.0006-0.006]$~\cite{Cline:2013gha}. This keeps
the invisible decay width of $h$ well within the observed limits.
Other model couplings remain unconstrained by these observations.

Within this simplistic framework, one expects the process $p p \to H \to h
\chi \chi$ to generate an enhanced $p_{T}$, owing to the fact that $h$
now recoils against a pair of invisible particles. The presence of a $H\chi\chi$ coupling opens the potential for detecting invisible decays following the methodology suggested in Ref.~\cite{Djouadi:2012zc}.

\section{Statistical formalism\label{sec:stats}}
The BSM prediction constructed using the proposed model was fit against four classes of constraints. The constraints considered are differential Higgs boson $p_T$ spectra, the limits on di-Higgs boson production through a resonance, various limits and measurements on top associated Higgs boson production, and the limits on a heavy scalar decaying to vector bosons. For each class of constraints, results from both ATLAS and CMS were used to avoid bias. These results are summarized in Table~\ref{tab:list_of_results}. A simultaneous fit was done in terms of calculating and minimising a combined $\chi^2$ value while varying $\beta_g$ for different $m_H$ hypotheses. Statistically, two types of results were dealt with: measurements and limits.

A calculation of $\chi^2$ for a measurement is straightforward. Given a measurement $\mu$ with its associated error $\Delta\mu$, one can construct a $\chi^2$ by testing the measurement against a theoretical prediction and its associated theoretical uncertainty, given as $\mu^{\text{th}}$ and $\Delta\mu^{\text{th}}$, respectively. The uncertainties from the measurement and the theoretical prediction are assumed to be independent and are added up in quadrature, allowing us to calculate the $\chi^2$ as

\begin{equation}
\chi^2=\frac{(\mu-\mu^{\text{th}})^2}{(\Delta\mu)^2+(\Delta\mu^{\text{th}})^2}.
\label{eqn:cs_meas}
\end{equation}

To calculate a $\chi^2$ from a result in the form of a 95\%~confidence limit (CL), we need only assume that given some measurement $\mu$ with its expected and observed limits $\mu^{\text{exp}}$ and $\mu^{\text{obs}}$ respectively, the $\chi^2$ is Gaussian in $\mu$. Then, assuming the null hypothesis has $\mu=0$, we can extract the mean of the distribution as $\mu^{\text{obs}}-\mu^{\text{exp}}$. This is treated as the excess in $\mu$ which can be tested against a theoretical prediction $\mu^{\text{th}}$. Its error is inserted as $\Delta\mu=\mu^{\text{exp}}/1.96$, where the 1.96 arises from the fact that 95\% confidence corresponds to 1.96 units of standard deviation in a Gaussian distribution -- this approach is used in Ref.~\cite{Giardino:2013bma} and other references therein. Using this, the $\chi^2$ is calculated as 

\begin{equation}
\chi^2=\frac{(\mu^{\text{obs}}-\mu^{\text{exp}}-\mu^{\text{th}})^2}{(\mu^{\text{exp}}/1.96)^2}.
\label{eqn:cs_lim}
\end{equation}

Using the definitions in Eqs.~(\ref{eqn:cs_meas}) and (\ref{eqn:cs_lim}), a combined $\chi^2$ was constructed by adding up the contributions from all of the results presented in Table~\ref{tab:list_of_results}. This procedure is described in the following section.

\section {Methodology and tools\label{sec:tools}}



\begin{figure*}
	\centering
	\begin{minipage}{0.48\textwidth}
		{\includegraphics[width=\textwidth]{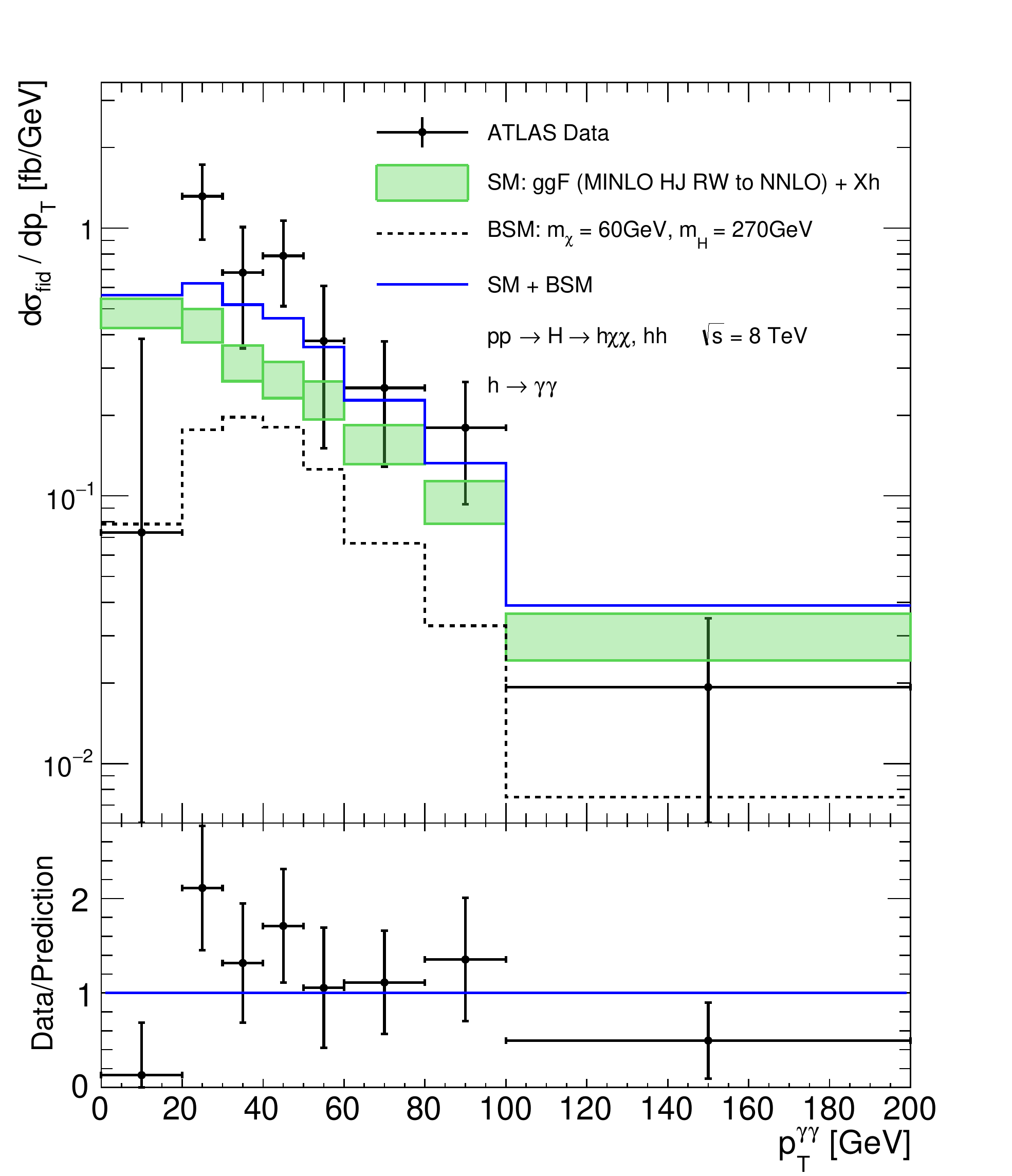}}
	\end{minipage}
	\begin{minipage}{0.49\textwidth}
		{\includegraphics[scale=0.4]{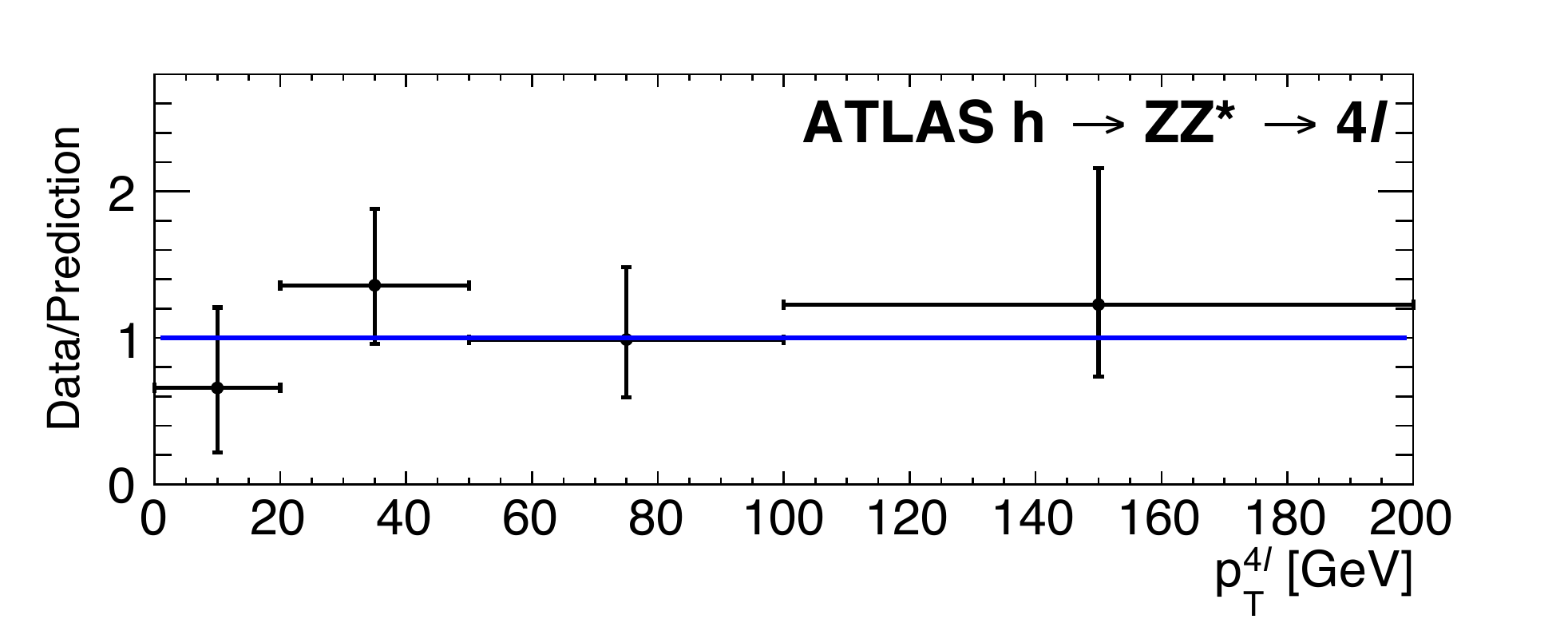}}
		{\includegraphics[scale=0.4]{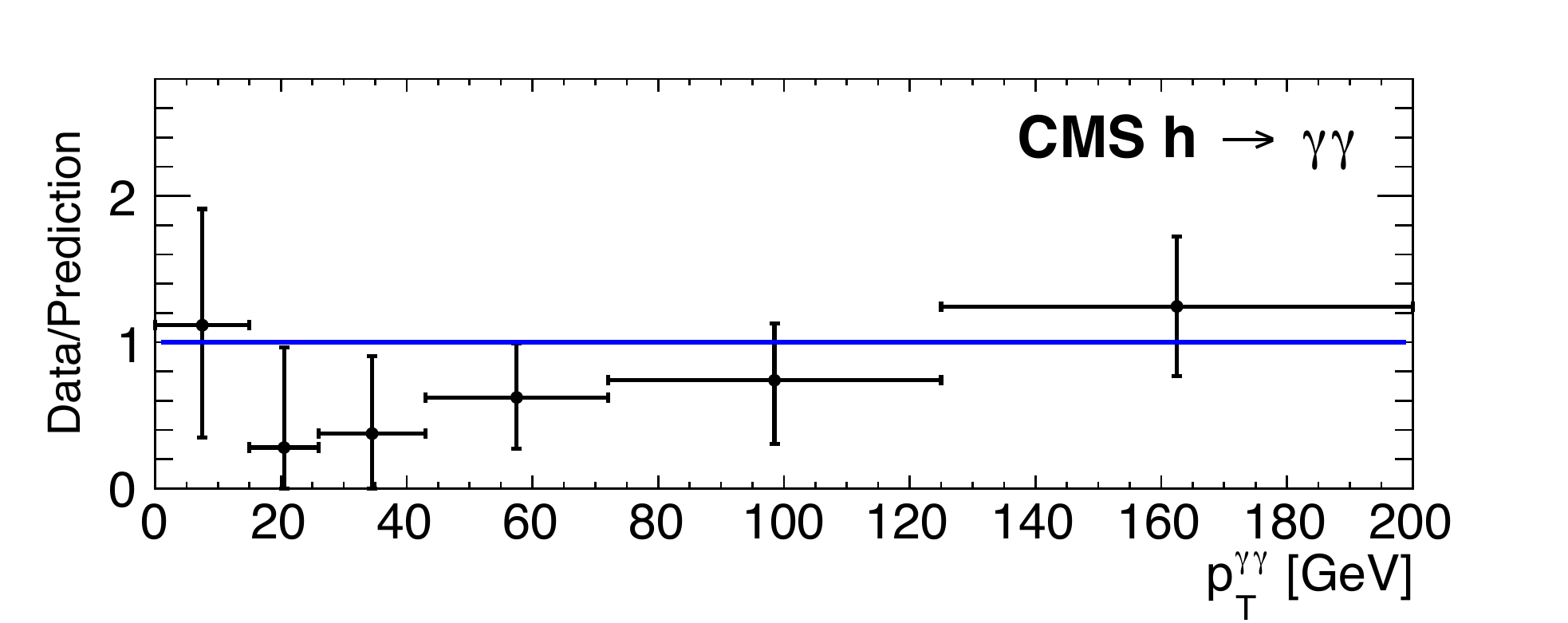}}
		{\includegraphics[scale=0.4]{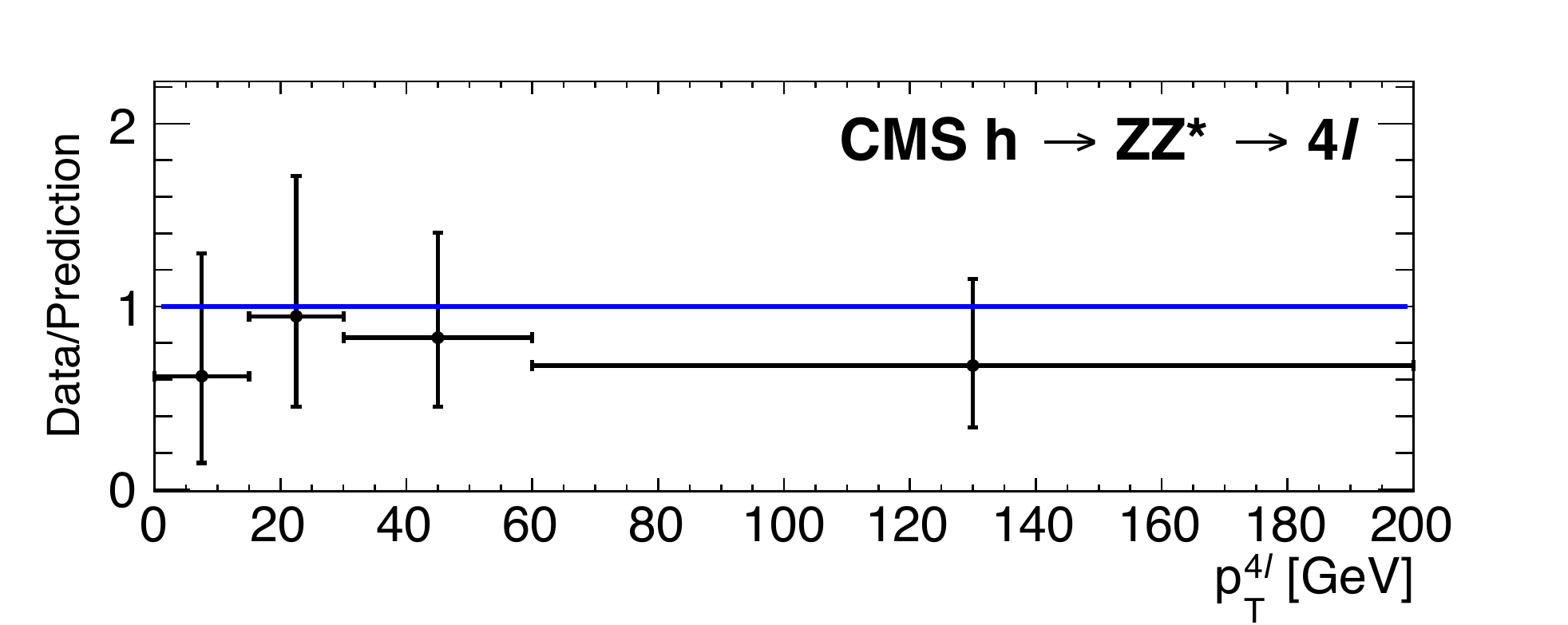}}
	\end{minipage}
	
	\caption{Fits to the fiducial differential distributions of the Higgs
		boson transverse momentum using the ATLAS diphoton~\cite{Aad:2014lwa}
		(left), the ATLAS $h\to ZZ^*\to 4\ell$~\cite{Aad:2014tca} (top right), the CMS diphoton~\cite{Khachatryan:2015rxa} (middle right) and CMS $h\to ZZ^*\to 4\ell$~\cite{CMS:2015hja} (bottom right)
		decays (see text for detailed description). The mass points
		considered here are the best fit values of $m\chi=60$~GeV and $m_H=270$~GeV. In the interest of saving space, only one full plot has been shown, with the other three are plots of a ratio between the data and BSM prediction.}
	\label{fig:fits}
\end{figure*}

The minimal model we have described was built first using
\texttt{FeynRules}~\cite{Alloul:2013bka} and then passed to the
Universal FeynRules Output~\cite{Degrande:2011ua} such that event
generation could be performed at leading order (LO) in \texttt{MadGraph5}~\cite{Alwall:2011uj}. Computations relating to DM
constraints in the model were carried out using
\texttt{micrOMEGAs}~\cite{Belanger:2013oya}.

Fitting the ATLAS and CMS Higgs boson $p_T$ spectra with the BSM prediction was accomplished as follows. Events with the $h\chi\chi$ and $hh$ final states were
generated from $pp$ collisions through an $H$ $s$-channel in \texttt{MadGraph5} and showered appropriately using
\texttt{PYTHIA 8.2}~\cite{Sjostrand:2014zea}. Since the ATLAS~\cite{Aad:2014lwa,Aad:2014tca} and CMS~\cite{Khachatryan:2015rxa,CMS:2015hja} Higgs boson $p_T$ spectra were constructed from fiducial volumes of phase space, it was important that BSM prediction went through the same event selection. This was done using the \texttt{Rivet}~\cite{Buckley:2010ar} analysis framework. The total LO cross
section of the BSM prediction was enhanced to NNLL+NLO (next-to-next-to leading log plus next-to leading order accuracy) through
multiplication by an appropriately calculated k-factor determined from
Ref.~\cite{Heinemeyer:2013tqa}.

The BSM prediction was tested against the SM-only prediction. For the SM, fiducial acceptance factors for gluon
fusion (ggF) are given in Refs. \cite{Aad:2014lwa,Aad:2014tca,Khachatryan:2015rxa,CMS:2015hja}. The ggF Higgs boson $p_T$ prediction was generated at NLO
using \texttt{MiNLO} HJ \cite{Hamilton:2012np}. This prediction was
further reweighted to NNLO accuracy and scaled to the NNLL+NLO cross section from
Ref.~\cite{Heinemeyer:2013tqa}. The other less prominent Higgs boson
production modes (VBF, $Vh$ and $t\bar{t}h$, which are collectively
referred to as $Xh$) were taken directly from the ATLAS and CMS
publications.

The bulk of the production cross section is in the intermediate
$p_{Th}$ range where ggF is the dominant production mode. In this
mechanism, QCD radiative corrections play a critical role in generating
$p_{Th}$. The Monte Carlo (MC) used to simulate ggF describes the $p_{Th}$
distribution at NNLL+NLO. Recent results on the NNLO corrections on
ggF+1j production indicate that, although moderate, corrections are
still significant~\cite{Boughezal:2015dra,Boughezal:2015aha}. NNLO corrections with
respect to NLO can be as large at 25\% in the range of interest. In
order to accommodate these corrections, a conservative approach is
implemented. The $p_{Th}$ distribution with $p_{Th}>30$~GeV is
corrected with the NNLO/NLO k-factors provided in
Ref.~\cite{Boughezal:2015dra}. The MC described is normalized to the
total ggF cross section at NNLO. The first complete calculation of the
total ggF cross section at N$^3$LO is now available and it is
indicative of small N$^3$LO/NNLO k-factors and scale
variations~\cite{Anastasiou:2015ema}. For this reason the
cross section with $p_{Th}<30$~GeV is re-scaled appropriately so that
the total cross section does not exceed the total cross section by
more than 2\% with respect to the calculation at NNLO. The scale
uncertainties assumed in this analysis remain at NNLO for the total
cross section and at NLO for the $p_{Th}$, while the PDF
uncertainties were conservatively taken from
Ref.~\cite{Heinemeyer:2013tqa}. Other production mechanisms of the SM
Higgs boson do not play a critical role in the region of the
phase-space under study, and were therefore left unmodified. A similar approach is taken for the BSM prediction. The $p_T$ of the heavy scalar is reweighted by determining a reweighting function from an NLO calculation using
\texttt{OneLOoP}~\cite{vanHameren:2010cp} in \texttt{MG5\_aMC@NLO}
\cite{Alwall:2014hca} compared to the result obtained with the
shower. Overall, the shower does a reasonable job, matching the LO
prediction within 20\%. The effect of the these corrections on the
transverse momentum of the Higgs boson from the decay of $H$
is small and it reduces to a positive shift of about 3~GeV. It
is, however, important to note that the jet multiplicity of the $H$
boson in this setup is significantly larger than that characteristic
to $h$. This implies a significant reduction of the jet veto survival
probability.

Excesses in top associated Higgs boson production were also included in the fits. Associated $th$ production in the SM is suppressed due to
the negative interference induced by the relative sign of the Yukawa
and $hWW$ couplings (see Ref.~\cite{Farina:2012xp} and other references
therein). If the $HWW$ coupling is relatively suppressed, this negative
interference is reduced, so
that its cross section becomes comparable to that of $ttH$ production. For this reason, $\beta_{_V}$ was set to a small value (order of $10^{-3}$) and $tH$ cross sections were determined at LO in \texttt{MadGraph5}. These cross sections were enhanced to NNLL+NLO by multiplying by an appropriate k-factor, and were then combined with $ttH$ cross sections from Ref.~\cite{Heinemeyer:2013tqa}. For the mass values of the heavy scalar considered in this analysis (between 260 and 320~GeV) the combined cross section from $tH$ and $ttH$ reached a value as high as 25~fb at $\sqrt{s}=8$~TeV.

The statistical combination was done using the techniques described in Section~\ref{sec:stats}. Firstly, the branching ratios of $H\to hh$ and $H\to VV$ were fixed by minimising a $\chi^2$ determined from experimental results. These branching ratios were used as inputs for a combined $\chi^2$, which was calculated while floating the free parameters $\beta_g$ and $m_H$. For each mass point, $\beta_g$ was marginalised such that the combined $\chi^2$ was minimised. Errors on marginalised parameters were calculated from identifying the points in parameter space which differ by one unit of $\chi^2$ above and below the minimised value.

\section{Results and discussion\label{sec:results}} 

In this analysis the global $\chi^2$ is minimized for different $m_H$ hypotheses. The technical part of the analysis was done using a scan of mass points, starting at $m_H=260$~GeV and going up in 5~GeV steps until 320~GeV. Points in between these were reached by an interpolation. The other parameters of the model are fixed by a number of
constraints. Firstly, the branching ratio of
$H\rightarrow h h$ is set to a value that is best fit against the current di-Higgs boson resonance search limits set by ATLAS and CMS. Secondly, the
branching ratio of $H\rightarrow VV$ is determined in the same way using ATLAS and CMS limits from searches for $H\to VV$ at high masses. The remainder of the decay of the heavy scalar is assumed to be $H\to h\chi\chi$. Finally, the parameter $\beta_g$ is constrained by fitting the ATLAS and CMS Higgs boson $p_T$ spectra, as well as excesses in top associated Higgs boson production. There may exist other decay modes, such as $H\to\chi\chi$. The invisible decay is not considered here. Adding in other decay modes would not change the final results of the analysis, although it would allow us to further constrain the parameter $\beta_g$.

Calculating and minimising the $\chi^2$ described in Sections~\ref{sec:stats} and \ref{sec:tools}, it was found that the lower values of $m_H$ fit the experimental data better than the higher values. Out of the mass points considered, the $m_H=270$~GeV point was able to minimise the $\chi^2$ value the most. This point was determined using the best fit values of the branching ratios as BR$(H\to hh)=0.030\pm0.037$,~BR$(H\to ZZ)=0.025\pm0.018$ and BR$(H\to WW)=0.057\pm0.041$. The parameter $\beta_g$ was best fit at the value of $1.5\pm0.6$. The errors on these quantities correspond to a $1\sigma$ deviation from the mean value. An indication of this parameter fitting the ATLAS and CMS $p_T$ spectra can be seen in Fig.~\ref{fig:fits}. The fits to the $p_T$ spectra were also able to constrain the mass of the DM candidate; for $m_H=270$~GeV, $m_\chi$ was best fit at 60~GeV. This is very close to $m_h/2$, which naturally leads to the suppression
of the branching ratio of $h\rightarrow \chi\chi$, and it is consistent
with current direct search limits. The distribution with the
filled band corresponds to the prediction made by the SM. The width of the band
indicates the size of the uncertainties on the ggF process,
according to the conservative scheme discussed in Section~\ref{sec:tools}. These uncertainties are incorporated in the $\chi^2$. The
dotted line shows the contribution from $H\rightarrow h \chi\chi$ as well as $H\to hh$. The solid line corresponds to the sum of the SM and BSM components. 

\begin{figure}[t]
	\centering
	\includegraphics[width=0.5\textwidth]{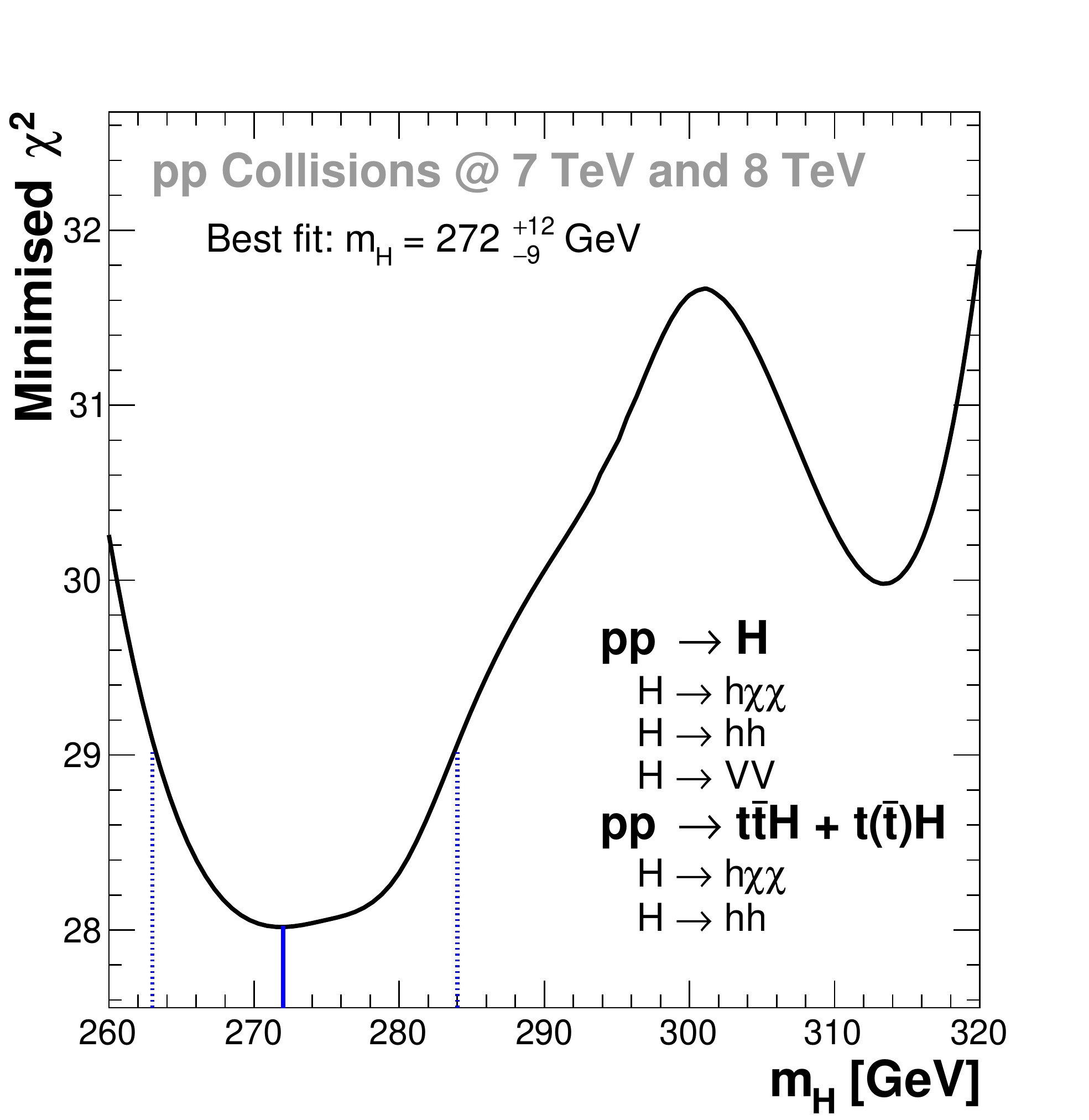}
	\caption{A scan of minimised $\chi^2$ values as a function of the free parameter $m_H$. This was constructed by minimising an additive $\chi^2$ with contributions from all of the experimental results in Table~\ref{tab:list_of_results}. These results were compared to the BSM prediction described mostly in Section~\ref{sec:tools}. }
	\label{fig:min_cs}
\end{figure}

\begin{figure}
	\centering
	\includegraphics[width=0.5\textwidth]{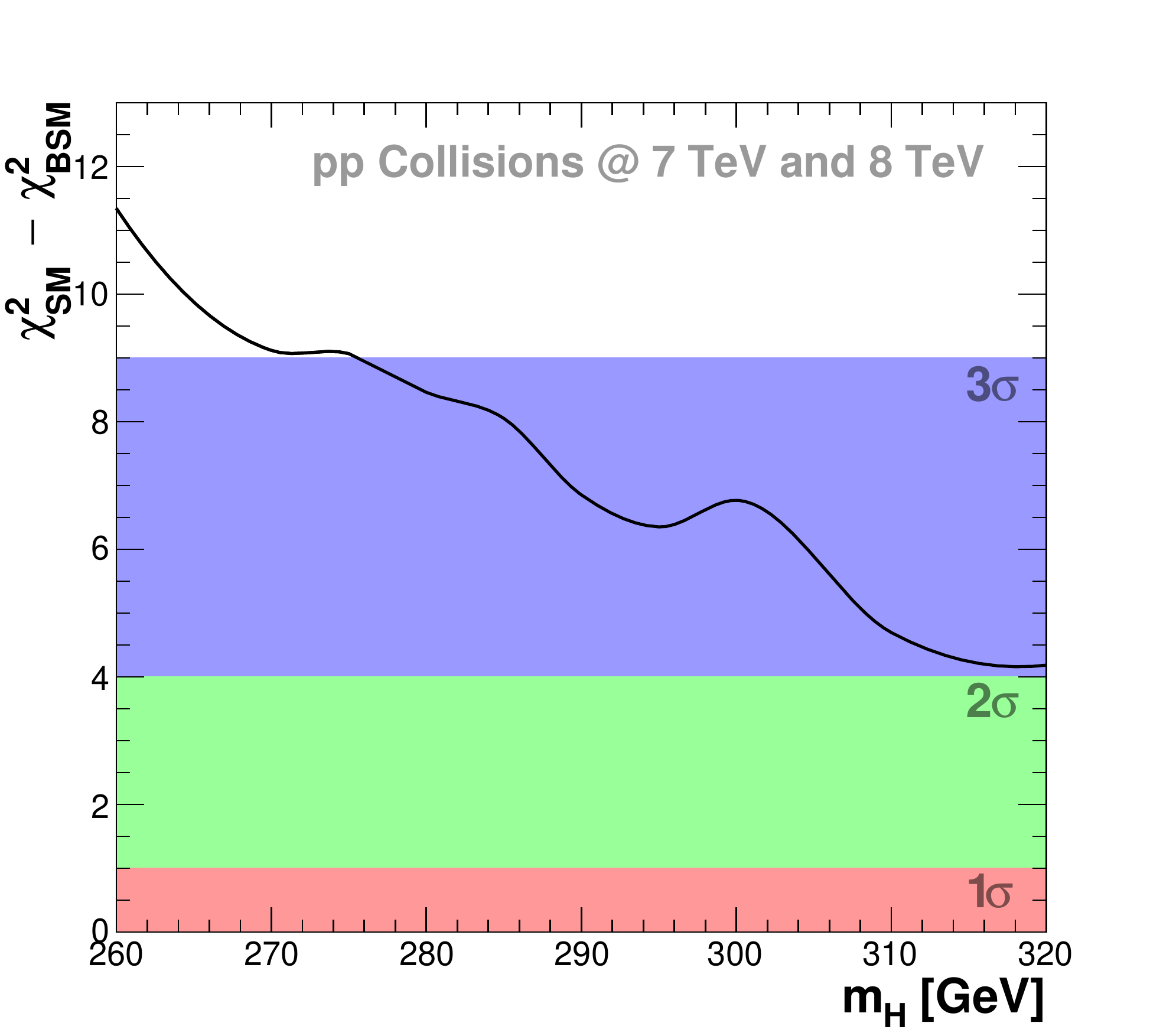}
	\caption{A scan over $m_H$ of the test statistic $\chi^2_{\text{SM}}-\chi^2_{\text{BSM}}$ employed for quantifying the significance of the effect of the BSM model. The $1\sigma$, $2\sigma$ and $3\sigma$ bands for one degree of freedom are shown in red, green and blue, respectively. The SM hypothesis corresponds to the existence of the SM Higgs boson, but the complete absence of the heavy scalar (i.e. the $pp\to H$ cross section is set to 0).}
	\label{fig:delta_cs}
\end{figure}

When interpolating between mass points, the combined minimised $\chi^2$ is found to be smallest at the value $m_H=272$~GeV, with upper and lower errors being 12~GeV and 9~GeV, respectively. This can be seen in Fig.~\ref{fig:min_cs} where the solid blue line shows the lowest value of the minimised $\chi^2$, and the dotted blue lines show a $1\sigma$ deviation from the value. The minimised value of $\chi^2$ has a lowest value of 0.72 per degree of freedom in the fit. It should be noted here that when comparing the BSM hypothesis to a null hypothesis (the SM with a 125~GeV Higgs boson), the improvement on explaining experimental data just surpasses a $3\sigma$ effect at the best fit point, as can be seen in Fig.~\ref{fig:delta_cs}. In this figure, the large significance around $m_H=260$~GeV can be attributed to the large $pp\to H\to hh$ cross sections in most of the ATLAS and CMS di-Higgs boson resonance search results. It is also relevant to note that
results reported here do not change significantly if the NNLO
corrections on ggF+1j discussed in the previous section are not
applied.

The most immediate consequence of the
phenomenological model considered here is the appearance of
intermediate missing transverse energy in association with $h$. In addition, data appears to display more jets in association to $h$ than expected in the SM. This applies both to the inclusive production and the production in association with top quarks. Enhanced QCD radiation in the production of $H$ compared to that of direct $h$ production may not be sufficient to explain the effect. As a result, one can also consider the decay of the hypothetical intermediate particle (discussed in Section~\ref{sec:intro}) into hadronic jets. These effects would lead to an enhanced jet multiplicity in the inclusive production of $h$ as well as in association with top quarks. This also opens the opportunity for a resonant structure in the $hjj$ spectrum.

In the light of this discussion, presented below is a list of possible smoking guns for investigation with Run 2 data:
\begin{itemize}
\item Higgs boson in association with moderate and large missing energy.
\item Higgs boson in association with at least two hadronic jets, including a resonant structure in the $hjj$ spectrum. 
\item Higgs boson in association with top quarks and large missing energy (greater than 100\,GeV)
\item Higgs boson in association with top quarks and at least two additional hadronic jets, including a resonant structure in the $hjj$ spectrum. 
\item Resonant structure in the $VV$ invariant mass spectrum.
\item Resonant structure in the $hh$ invariant mass spectrum.
\item Missing energy recoiling against a high transverse momentum jet.
\end{itemize}

It is beyond the scope of this paper to quantify the amount of integrated luminosity required for the ATLAS and CMS experiments to declare a 5$\sigma$ effect. However, it is noted that the design of dedicated data analysis will greatly enhance the sensitivity reached with the available results from Run 1 studied here. 

\section{Conclusions\label{sec:conclusions}} 

ATLAS and CMS data results comprising measurement of the differential Higgs boson transverse momentum, searches for a di-Higgs boson resonance, the Higgs boson in association with top quarks, and $VV$ resonances, have been considered against the hypothesis of a heavy scalar. The analysis yields a best fit result with a 3$\sigma$ effect. The hypothetical heavy boson mass is measured to be 272$^{+12}_{-9}$\,GeV. In this setup the heavy scalar would predominantly decay into at least one Higgs boson in association with missing energy or hadronic jets. These results are obtained on the basis of analyses that are not optimized for the search of the heavy scalar discussed here. It is expected that dedicated optimizations will significantly enhance the sensitivity to the search. A number of final states have been identified that can serve as smoking guns for the discovery of the heavy scalar discussed here. 

\section*{Acknowledgements} 

The work of N.C., T.M. and B. Mukhopadhyaya
was partially supported by funding available from the Department of
Atomic Energy, Government of India for the Regional Centre for
Accelerator-based Particle Physics (RECAPP), Harish-Chandra Research
Institute. B. Mellado acknowledges the hospitality of RECAPP during
the collaboration. The Claude Leon
Foundation are acknowledged for their financial support. The High
Energy Physics group of the University of the Witwatersrand is
grateful for the support from the Wits Research Office, the National
Research Foundation, the National Institute of Theoretical Physics and
the Department of Science and Technology through the SA-CERN
consortium and other forms of support.

\section*{References}

\bibliographystyle{apsrev4-1}
\bibliography{ref}

\begin{thebibliography}{39}%
\makeatletter
\providecommand \@ifxundefined [1]{%
 \@ifx{#1\undefined}
}%
\providecommand \@ifnum [1]{%
 \ifnum #1\expandafter \@firstoftwo
 \else \expandafter \@secondoftwo
 \fi
}%
\providecommand \@ifx [1]{%
 \ifx #1\expandafter \@firstoftwo
 \else \expandafter \@secondoftwo
 \fi
}%
\providecommand \natexlab [1]{#1}%
\providecommand \enquote  [1]{``#1''}%
\providecommand \bibnamefont  [1]{#1}%
\providecommand \bibfnamefont [1]{#1}%
\providecommand \citenamefont [1]{#1}%
\providecommand \href@noop [0]{\@secondoftwo}%
\providecommand \href [0]{\begingroup \@sanitize@url \@href}%
\providecommand \@href[1]{\@@startlink{#1}\@@href}%
\providecommand \@@href[1]{\endgroup#1\@@endlink}%
\providecommand \@sanitize@url [0]{\catcode `\\12\catcode `\$12\catcode
  `\&12\catcode `\#12\catcode `\^12\catcode `\_12\catcode `\%12\relax}%
\providecommand \@@startlink[1]{}%
\providecommand \@@endlink[0]{}%
\providecommand \url  [0]{\begingroup\@sanitize@url \@url }%
\providecommand \@url [1]{\endgroup\@href {#1}{\urlprefix }}%
\providecommand \urlprefix  [0]{URL }%
\providecommand \Eprint [0]{\href }%
\providecommand \doibase [0]{http://dx.doi.org/}%
\providecommand \selectlanguage [0]{\@gobble}%
\providecommand \bibinfo  [0]{\@secondoftwo}%
\providecommand \bibfield  [0]{\@secondoftwo}%
\providecommand \translation [1]{[#1]}%
\providecommand \BibitemOpen [0]{}%
\providecommand \bibitemStop [0]{}%
\providecommand \bibitemNoStop [0]{.\EOS\space}%
\providecommand \EOS [0]{\spacefactor3000\relax}%
\providecommand \BibitemShut  [1]{\csname bibitem#1\endcsname}%
\let\auto@bib@innerbib\@empty
\bibitem [{\citenamefont {Aad}\ \emph {et~al.}(2013)\citenamefont {Aad} \emph
  {et~al.}}]{Aad:2012tfa}%
  \BibitemOpen
  \bibfield  {author} {\bibinfo {author} {\bibfnamefont {G.}~\bibnamefont
  {Aad}} \emph {et~al.} (\bibinfo {collaboration} {ATLAS Collaboration}),\
  }\href {\doibase 10.1016/j.physletb.2012.08.020} {\bibfield  {journal}
  {\bibinfo  {journal} {Phys. Lett.}\ }\textbf {\bibinfo {volume} {B716}},\
  \bibinfo {pages} {1} (\bibinfo {year} {2013})},\ \Eprint
  {http://arxiv.org/abs/1207.7214} {arXiv:1207.7214 [hep-ex]} \BibitemShut
  {NoStop}%
\bibitem [{\citenamefont {Chatrchyan}\ \emph {et~al.}(2012)\citenamefont
  {Chatrchyan} \emph {et~al.}}]{Chatrchyan:2012xdj}%
  \BibitemOpen
  \bibfield  {author} {\bibinfo {author} {\bibfnamefont {S.}~\bibnamefont
  {Chatrchyan}} \emph {et~al.} (\bibinfo {collaboration} {CMS Collaboration}),\
  }\href {\doibase 10.1016/j.physletb.2012.08.021} {\bibfield  {journal}
  {\bibinfo  {journal} {Phys. Lett.}\ }\textbf {\bibinfo {volume} {B716}},\
  \bibinfo {pages} {30} (\bibinfo {year} {2012})},\ \Eprint
  {http://arxiv.org/abs/1207.7235} {arXiv:1207.7235 [hep-ex]} \BibitemShut
  {NoStop}%
\bibitem [{\citenamefont {Aad}\ \emph {et~al.}(2014{\natexlab{a}})\citenamefont
  {Aad} \emph {et~al.}}]{Aad:2014lwa}%
  \BibitemOpen
  \bibfield  {author} {\bibinfo {author} {\bibfnamefont {G.}~\bibnamefont
  {Aad}} \emph {et~al.} (\bibinfo {collaboration} {ATLAS Collaboration}),\
  }\href {\doibase 10.1007/JHEP09(2014)112} {\bibfield  {journal} {\bibinfo
  {journal} {JHEP}\ }\textbf {\bibinfo {volume} {09}},\ \bibinfo {pages} {112}
  (\bibinfo {year} {2014}{\natexlab{a}})},\ \Eprint
  {http://arxiv.org/abs/1407.4222} {arXiv:1407.4222 [hep-ex]} \BibitemShut
  {NoStop}%
\bibitem [{\citenamefont {Aad}\ \emph {et~al.}(2014{\natexlab{b}})\citenamefont
  {Aad} \emph {et~al.}}]{Aad:2014tca}%
  \BibitemOpen
  \bibfield  {author} {\bibinfo {author} {\bibfnamefont {G.}~\bibnamefont
  {Aad}} \emph {et~al.} (\bibinfo {collaboration} {ATLAS Collaboration}),\
  }\href {\doibase 10.1016/j.physletb.2014.09.054} {\bibfield  {journal}
  {\bibinfo  {journal} {Phys. Lett.}\ }\textbf {\bibinfo {volume} {B738}},\
  \bibinfo {pages} {234} (\bibinfo {year} {2014}{\natexlab{b}})},\ \Eprint
  {http://arxiv.org/abs/1408.3226} {arXiv:1408.3226 [hep-ex]} \BibitemShut
  {NoStop}%
\bibitem [{\citenamefont {Khachatryan}\ \emph
  {et~al.}(2015{\natexlab{a}})\citenamefont {Khachatryan} \emph
  {et~al.}}]{Khachatryan:2015rxa}%
  \BibitemOpen
  \bibfield  {author} {\bibinfo {author} {\bibfnamefont {V.}~\bibnamefont
  {Khachatryan}} \emph {et~al.} (\bibinfo {collaboration} {CMS
  Collaboration}),\ }\href@noop {} {\  (\bibinfo {year}
  {2015}{\natexlab{a}})},\ \Eprint {http://arxiv.org/abs/1508.07819}
  {arXiv:1508.07819 [hep-ex]} \BibitemShut {NoStop}%
\bibitem [{\citenamefont {{CMS Collaboration}}(2015)}]{CMS:2015hja}%
  \BibitemOpen
  \bibfield  {author} {\bibinfo {author} {\bibnamefont {{CMS Collaboration}}},\
  }\href@noop {} {\  (\bibinfo {year} {2015})},\ \bibinfo {note}
  {{CMS-PAS-HIG-14-028}}\BibitemShut {NoStop}%
\bibitem [{\citenamefont {Aad}\ \emph {et~al.}(2015{\natexlab{a}})\citenamefont
  {Aad} \emph {et~al.}}]{Aad:2015xja}%
  \BibitemOpen
  \bibfield  {author} {\bibinfo {author} {\bibfnamefont {G.}~\bibnamefont
  {Aad}} \emph {et~al.} (\bibinfo {collaboration} {ATLAS Collaboration}),\
  }\href@noop {} {\  (\bibinfo {year} {2015}{\natexlab{a}})},\ \Eprint
  {http://arxiv.org/abs/1509.04670} {arXiv:1509.04670 [hep-ex]} \BibitemShut
  {NoStop}%
\bibitem [{\citenamefont {{CMS Collaboration}}(2014)}]{CMS:2014ipa}%
  \BibitemOpen
  \bibfield  {author} {\bibinfo {author} {\bibnamefont {{CMS Collaboration}}},\
  }\href@noop {} {\  (\bibinfo {year} {2014})},\ \bibinfo {note}
  {{CMS-PAS-HIG-13-032}}\BibitemShut {NoStop}%
\bibitem [{\citenamefont {Khachatryan}\ \emph
  {et~al.}(2015{\natexlab{b}})\citenamefont {Khachatryan} \emph
  {et~al.}}]{Khachatryan:2015tha}%
  \BibitemOpen
  \bibfield  {author} {\bibinfo {author} {\bibfnamefont {V.}~\bibnamefont
  {Khachatryan}} \emph {et~al.} (\bibinfo {collaboration} {CMS
  Collaboration}),\ }\href@noop {} {\  (\bibinfo {year}
  {2015}{\natexlab{b}})},\ \Eprint {http://arxiv.org/abs/1510.01181}
  {arXiv:1510.01181 [hep-ex]} \BibitemShut {NoStop}%
\bibitem [{\citenamefont {Khachatryan}\ \emph
  {et~al.}(2014{\natexlab{a}})\citenamefont {Khachatryan} \emph
  {et~al.}}]{Khachatryan:2014jya}%
  \BibitemOpen
  \bibfield  {author} {\bibinfo {author} {\bibfnamefont {V.}~\bibnamefont
  {Khachatryan}} \emph {et~al.} (\bibinfo {collaboration} {CMS
  Collaboration}),\ }\href {\doibase 10.1103/PhysRevD.90.112013} {\bibfield
  {journal} {\bibinfo  {journal} {Phys. Rev.}\ }\textbf {\bibinfo {volume}
  {D90}},\ \bibinfo {pages} {112013} (\bibinfo {year} {2014}{\natexlab{a}})},\
  \Eprint {http://arxiv.org/abs/1410.2751} {arXiv:1410.2751 [hep-ex]}
  \BibitemShut {NoStop}%
\bibitem [{\citenamefont {Aad}\ \emph {et~al.}(2015{\natexlab{b}})\citenamefont
  {Aad} \emph {et~al.}}]{Aad:2014lma}%
  \BibitemOpen
  \bibfield  {author} {\bibinfo {author} {\bibfnamefont {G.}~\bibnamefont
  {Aad}} \emph {et~al.} (\bibinfo {collaboration} {ATLAS Collaboration}),\
  }\href {\doibase 10.1016/j.physletb.2014.11.049} {\bibfield  {journal}
  {\bibinfo  {journal} {Phys. Lett.}\ }\textbf {\bibinfo {volume} {B740}},\
  \bibinfo {pages} {222} (\bibinfo {year} {2015}{\natexlab{b}})},\ \Eprint
  {http://arxiv.org/abs/1409.3122} {arXiv:1409.3122 [hep-ex]} \BibitemShut
  {NoStop}%
\bibitem [{\citenamefont {Aad}\ \emph {et~al.}(2015{\natexlab{c}})\citenamefont
  {Aad} \emph {et~al.}}]{Aad:2015iha}%
  \BibitemOpen
  \bibfield  {author} {\bibinfo {author} {\bibfnamefont {G.}~\bibnamefont
  {Aad}} \emph {et~al.} (\bibinfo {collaboration} {ATLAS Collaboration}),\
  }\href {\doibase 10.1016/j.physletb.2015.07.079} {\bibfield  {journal}
  {\bibinfo  {journal} {Phys. Lett.}\ }\textbf {\bibinfo {volume} {B749}},\
  \bibinfo {pages} {519} (\bibinfo {year} {2015}{\natexlab{c}})},\ \Eprint
  {http://arxiv.org/abs/1506.05988} {arXiv:1506.05988 [hep-ex]} \BibitemShut
  {NoStop}%
\bibitem [{\citenamefont {Aad}\ \emph {et~al.}(2015{\natexlab{d}})\citenamefont
  {Aad} \emph {et~al.}}]{Aad:2015gra}%
  \BibitemOpen
  \bibfield  {author} {\bibinfo {author} {\bibfnamefont {G.}~\bibnamefont
  {Aad}} \emph {et~al.} (\bibinfo {collaboration} {ATLAS Collaboration}),\
  }\href {\doibase 10.1140/epjc/s10052-015-3543-1} {\bibfield  {journal}
  {\bibinfo  {journal} {Eur. Phys. J.}\ }\textbf {\bibinfo {volume} {C75}},\
  \bibinfo {pages} {349} (\bibinfo {year} {2015}{\natexlab{d}})},\ \Eprint
  {http://arxiv.org/abs/1503.05066} {arXiv:1503.05066 [hep-ex]} \BibitemShut
  {NoStop}%
\bibitem [{\citenamefont {Khachatryan}\ \emph
  {et~al.}(2014{\natexlab{b}})\citenamefont {Khachatryan} \emph
  {et~al.}}]{Khachatryan:2014qaa}%
  \BibitemOpen
  \bibfield  {author} {\bibinfo {author} {\bibfnamefont {V.}~\bibnamefont
  {Khachatryan}} \emph {et~al.} (\bibinfo {collaboration} {CMS
  Collaboration}),\ }\href {\doibase 10.1007/JHEP09(2014)087,
  10.1007/JHEP10(2014)106} {\bibfield  {journal} {\bibinfo  {journal} {JHEP}\
  }\textbf {\bibinfo {volume} {09}},\ \bibinfo {pages} {087} (\bibinfo {year}
  {2014}{\natexlab{b}})},\ \bibinfo {note} {[Erratum: JHEP10,106(2014)]},\
  \Eprint {http://arxiv.org/abs/1408.1682} {arXiv:1408.1682 [hep-ex]}
  \BibitemShut {NoStop}%
\bibitem [{\citenamefont {Aad}\ \emph {et~al.}(2015{\natexlab{e}})\citenamefont
  {Aad} \emph {et~al.}}]{Aad:2015agg}%
  \BibitemOpen
  \bibfield  {author} {\bibinfo {author} {\bibfnamefont {G.}~\bibnamefont
  {Aad}} \emph {et~al.} (\bibinfo {collaboration} {ATLAS Collaboration}),\
  }\href@noop {} {\  (\bibinfo {year} {2015}{\natexlab{e}})},\ \Eprint
  {http://arxiv.org/abs/1509.00389} {arXiv:1509.00389 [hep-ex]} \BibitemShut
  {NoStop}%
\bibitem [{\citenamefont {Aad}\ \emph {et~al.}(2015{\natexlab{f}})\citenamefont
  {Aad} \emph {et~al.}}]{Aad:2015kna}%
  \BibitemOpen
  \bibfield  {author} {\bibinfo {author} {\bibfnamefont {G.}~\bibnamefont
  {Aad}} \emph {et~al.} (\bibinfo {collaboration} {ATLAS Collaboration}),\
  }\href@noop {} {\  (\bibinfo {year} {2015}{\natexlab{f}})},\ \Eprint
  {http://arxiv.org/abs/1507.05930} {arXiv:1507.05930 [hep-ex]} \BibitemShut
  {NoStop}%
\bibitem [{\citenamefont {Khachatryan}\ \emph
  {et~al.}(2015{\natexlab{c}})\citenamefont {Khachatryan} \emph
  {et~al.}}]{Khachatryan:2015cwa}%
  \BibitemOpen
  \bibfield  {author} {\bibinfo {author} {\bibfnamefont {V.}~\bibnamefont
  {Khachatryan}} \emph {et~al.} (\bibinfo {collaboration} {CMS
  Collaboration}),\ }\href@noop {} {\  (\bibinfo {year}
  {2015}{\natexlab{c}})},\ \Eprint {http://arxiv.org/abs/1504.00936}
  {arXiv:1504.00936 [hep-ex]} \BibitemShut {NoStop}%
\bibitem [{\citenamefont {Guo}\ and\ \citenamefont {Wu}(2010)}]{Guo:2010hq}%
  \BibitemOpen
  \bibfield  {author} {\bibinfo {author} {\bibfnamefont {W.-L.}\ \bibnamefont
  {Guo}}\ and\ \bibinfo {author} {\bibfnamefont {Y.-L.}\ \bibnamefont {Wu}},\
  }\href {\doibase 10.1007/JHEP10(2010)083} {\bibfield  {journal} {\bibinfo
  {journal} {JHEP}\ }\textbf {\bibinfo {volume} {10}},\ \bibinfo {pages} {083}
  (\bibinfo {year} {2010})},\ \Eprint {http://arxiv.org/abs/1006.2518}
  {arXiv:1006.2518 [hep-ph]} \BibitemShut {NoStop}%
\bibitem [{\citenamefont {Cai}\ \emph {et~al.}(2011)\citenamefont {Cai},
  \citenamefont {He},\ and\ \citenamefont {Ren}}]{Cai:2011kb}%
  \BibitemOpen
  \bibfield  {author} {\bibinfo {author} {\bibfnamefont {Y.}~\bibnamefont
  {Cai}}, \bibinfo {author} {\bibfnamefont {X.-G.}\ \bibnamefont {He}}, \ and\
  \bibinfo {author} {\bibfnamefont {B.}~\bibnamefont {Ren}},\ }\href {\doibase
  10.1103/PhysRevD.83.083524} {\bibfield  {journal} {\bibinfo  {journal} {Phys.
  Rev.}\ }\textbf {\bibinfo {volume} {D83}},\ \bibinfo {pages} {083524}
  (\bibinfo {year} {2011})},\ \Eprint {http://arxiv.org/abs/1102.1522}
  {arXiv:1102.1522 [hep-ph]} \BibitemShut {NoStop}%
\bibitem [{\citenamefont {Gonderinger}\ \emph {et~al.}(2012)\citenamefont
  {Gonderinger}, \citenamefont {Lim},\ and\ \citenamefont
  {Ramsey-Musolf}}]{Gonderinger:2012rd}%
  \BibitemOpen
  \bibfield  {author} {\bibinfo {author} {\bibfnamefont {M.}~\bibnamefont
  {Gonderinger}}, \bibinfo {author} {\bibfnamefont {H.}~\bibnamefont {Lim}}, \
  and\ \bibinfo {author} {\bibfnamefont {M.~J.}\ \bibnamefont
  {Ramsey-Musolf}},\ }\href {\doibase 10.1103/PhysRevD.86.043511} {\bibfield
  {journal} {\bibinfo  {journal} {Phys. Rev.}\ }\textbf {\bibinfo {volume}
  {D86}},\ \bibinfo {pages} {043511} (\bibinfo {year} {2012})},\ \Eprint
  {http://arxiv.org/abs/1202.1316} {arXiv:1202.1316 [hep-ph]} \BibitemShut
  {NoStop}%
\bibitem [{\citenamefont {Ade}\ \emph {et~al.}(2014)\citenamefont {Ade} \emph
  {et~al.}}]{Ade:2013zuv}%
  \BibitemOpen
  \bibfield  {author} {\bibinfo {author} {\bibfnamefont {P.~A.~R.}\
  \bibnamefont {Ade}} \emph {et~al.} (\bibinfo {collaboration} {Planck}),\
  }\href {\doibase 10.1051/0004-6361/201321591} {\bibfield  {journal} {\bibinfo
   {journal} {Astron. Astrophys.}\ }\textbf {\bibinfo {volume} {571}},\
  \bibinfo {pages} {A16} (\bibinfo {year} {2014})},\ \Eprint
  {http://arxiv.org/abs/1303.5076} {arXiv:1303.5076 [astro-ph.CO]} \BibitemShut
  {NoStop}%
\bibitem [{\citenamefont {Akerib}\ \emph {et~al.}(2014)\citenamefont {Akerib}
  \emph {et~al.}}]{Akerib:2013tjd}%
  \BibitemOpen
  \bibfield  {author} {\bibinfo {author} {\bibfnamefont {D.~S.}\ \bibnamefont
  {Akerib}} \emph {et~al.} (\bibinfo {collaboration} {LUX}),\ }\href {\doibase
  10.1103/PhysRevLett.112.091303} {\bibfield  {journal} {\bibinfo  {journal}
  {Phys. Rev. Lett.}\ }\textbf {\bibinfo {volume} {112}},\ \bibinfo {pages}
  {091303} (\bibinfo {year} {2014})},\ \Eprint {http://arxiv.org/abs/1310.8214}
  {arXiv:1310.8214 [astro-ph.CO]} \BibitemShut {NoStop}%
\bibitem [{\citenamefont {Cline}\ \emph {et~al.}(2013)\citenamefont {Cline},
  \citenamefont {Kainulainen}, \citenamefont {Scott},\ and\ \citenamefont
  {Weniger}}]{Cline:2013gha}%
  \BibitemOpen
  \bibfield  {author} {\bibinfo {author} {\bibfnamefont {J.~M.}\ \bibnamefont
  {Cline}}, \bibinfo {author} {\bibfnamefont {K.}~\bibnamefont {Kainulainen}},
  \bibinfo {author} {\bibfnamefont {P.}~\bibnamefont {Scott}}, \ and\ \bibinfo
  {author} {\bibfnamefont {C.}~\bibnamefont {Weniger}},\ }\href {\doibase
  10.1103/PhysRevD.92.039906, 10.1103/PhysRevD.88.055025} {\bibfield  {journal}
  {\bibinfo  {journal} {Phys. Rev.}\ }\textbf {\bibinfo {volume} {D88}},\
  \bibinfo {pages} {055025} (\bibinfo {year} {2013})},\ \bibinfo {note}
  {[Erratum: Phys. Rev.D92,no.3,039906(2015)]},\ \Eprint
  {http://arxiv.org/abs/1306.4710} {arXiv:1306.4710 [hep-ph]} \BibitemShut
  {NoStop}%
\bibitem [{\citenamefont {Djouadi}\ \emph {et~al.}(2013)\citenamefont
  {Djouadi}, \citenamefont {Falkowski}, \citenamefont {Mambrini},\ and\
  \citenamefont {Quevillon}}]{Djouadi:2012zc}%
  \BibitemOpen
  \bibfield  {author} {\bibinfo {author} {\bibfnamefont {A.}~\bibnamefont
  {Djouadi}}, \bibinfo {author} {\bibfnamefont {A.}~\bibnamefont {Falkowski}},
  \bibinfo {author} {\bibfnamefont {Y.}~\bibnamefont {Mambrini}}, \ and\
  \bibinfo {author} {\bibfnamefont {J.}~\bibnamefont {Quevillon}},\ }\href
  {\doibase 10.1140/epjc/s10052-013-2455-1} {\bibfield  {journal} {\bibinfo
  {journal} {Eur. Phys. J.}\ }\textbf {\bibinfo {volume} {C73}},\ \bibinfo
  {pages} {2455} (\bibinfo {year} {2013})},\ \Eprint
  {http://arxiv.org/abs/1205.3169} {arXiv:1205.3169 [hep-ph]} \BibitemShut
  {NoStop}%
\bibitem [{\citenamefont {Giardino}\ \emph {et~al.}(2014)\citenamefont
  {Giardino}, \citenamefont {Kannike}, \citenamefont {Masina}, \citenamefont
  {Raidal},\ and\ \citenamefont {Strumia}}]{Giardino:2013bma}%
  \BibitemOpen
  \bibfield  {author} {\bibinfo {author} {\bibfnamefont {P.~P.}\ \bibnamefont
  {Giardino}}, \bibinfo {author} {\bibfnamefont {K.}~\bibnamefont {Kannike}},
  \bibinfo {author} {\bibfnamefont {I.}~\bibnamefont {Masina}}, \bibinfo
  {author} {\bibfnamefont {M.}~\bibnamefont {Raidal}}, \ and\ \bibinfo {author}
  {\bibfnamefont {A.}~\bibnamefont {Strumia}},\ }\href {\doibase
  10.1007/JHEP05(2014)046} {\bibfield  {journal} {\bibinfo  {journal} {JHEP}\
  }\textbf {\bibinfo {volume} {05}},\ \bibinfo {pages} {046} (\bibinfo {year}
  {2014})},\ \Eprint {http://arxiv.org/abs/1303.3570} {arXiv:1303.3570
  [hep-ph]} \BibitemShut {NoStop}%
\bibitem [{\citenamefont {Alloul}\ \emph {et~al.}(2014)\citenamefont {Alloul},
  \citenamefont {Christensen}, \citenamefont {Degrande}, \citenamefont {Duhr},\
  and\ \citenamefont {Fuks}}]{Alloul:2013bka}%
  \BibitemOpen
  \bibfield  {author} {\bibinfo {author} {\bibfnamefont {A.}~\bibnamefont
  {Alloul}}, \bibinfo {author} {\bibfnamefont {N.~D.}\ \bibnamefont
  {Christensen}}, \bibinfo {author} {\bibfnamefont {C.}~\bibnamefont
  {Degrande}}, \bibinfo {author} {\bibfnamefont {C.}~\bibnamefont {Duhr}}, \
  and\ \bibinfo {author} {\bibfnamefont {B.}~\bibnamefont {Fuks}},\ }\href
  {\doibase 10.1016/j.cpc.2014.04.012} {\bibfield  {journal} {\bibinfo
  {journal} {Comput. Phys. Commun.}\ }\textbf {\bibinfo {volume} {185}},\
  \bibinfo {pages} {2250} (\bibinfo {year} {2014})},\ \Eprint
  {http://arxiv.org/abs/1310.1921} {arXiv:1310.1921 [hep-ph]} \BibitemShut
  {NoStop}%
\bibitem [{\citenamefont {Degrande}\ \emph {et~al.}(2012)\citenamefont
  {Degrande}, \citenamefont {Duhr}, \citenamefont {Fuks}, \citenamefont
  {Grellscheid}, \citenamefont {Mattelaer},\ and\ \citenamefont
  {Reiter}}]{Degrande:2011ua}%
  \BibitemOpen
  \bibfield  {author} {\bibinfo {author} {\bibfnamefont {C.}~\bibnamefont
  {Degrande}}, \bibinfo {author} {\bibfnamefont {C.}~\bibnamefont {Duhr}},
  \bibinfo {author} {\bibfnamefont {B.}~\bibnamefont {Fuks}}, \bibinfo {author}
  {\bibfnamefont {D.}~\bibnamefont {Grellscheid}}, \bibinfo {author}
  {\bibfnamefont {O.}~\bibnamefont {Mattelaer}}, \ and\ \bibinfo {author}
  {\bibfnamefont {T.}~\bibnamefont {Reiter}},\ }\href {\doibase
  10.1016/j.cpc.2012.01.022} {\bibfield  {journal} {\bibinfo  {journal}
  {Comput. Phys. Commun.}\ }\textbf {\bibinfo {volume} {183}},\ \bibinfo
  {pages} {1201} (\bibinfo {year} {2012})},\ \Eprint
  {http://arxiv.org/abs/1108.2040} {arXiv:1108.2040 [hep-ph]} \BibitemShut
  {NoStop}%
\bibitem [{\citenamefont {Alwall}\ \emph {et~al.}(2011)\citenamefont {Alwall},
  \citenamefont {Herquet}, \citenamefont {Maltoni}, \citenamefont {Mattelaer},\
  and\ \citenamefont {Stelzer}}]{Alwall:2011uj}%
  \BibitemOpen
  \bibfield  {author} {\bibinfo {author} {\bibfnamefont {J.}~\bibnamefont
  {Alwall}}, \bibinfo {author} {\bibfnamefont {M.}~\bibnamefont {Herquet}},
  \bibinfo {author} {\bibfnamefont {F.}~\bibnamefont {Maltoni}}, \bibinfo
  {author} {\bibfnamefont {O.}~\bibnamefont {Mattelaer}}, \ and\ \bibinfo
  {author} {\bibfnamefont {T.}~\bibnamefont {Stelzer}},\ }\href {\doibase
  10.1007/JHEP06(2011)128} {\bibfield  {journal} {\bibinfo  {journal} {JHEP}\
  }\textbf {\bibinfo {volume} {06}},\ \bibinfo {pages} {128} (\bibinfo {year}
  {2011})},\ \Eprint {http://arxiv.org/abs/1106.0522} {arXiv:1106.0522
  [hep-ph]} \BibitemShut {NoStop}%
\bibitem [{\citenamefont {Belanger}\ \emph {et~al.}(2014)\citenamefont
  {Belanger}, \citenamefont {Boudjema}, \citenamefont {Pukhov},\ and\
  \citenamefont {Semenov}}]{Belanger:2013oya}%
  \BibitemOpen
  \bibfield  {author} {\bibinfo {author} {\bibfnamefont {G.}~\bibnamefont
  {Belanger}}, \bibinfo {author} {\bibfnamefont {F.}~\bibnamefont {Boudjema}},
  \bibinfo {author} {\bibfnamefont {A.}~\bibnamefont {Pukhov}}, \ and\ \bibinfo
  {author} {\bibfnamefont {A.}~\bibnamefont {Semenov}},\ }\href {\doibase
  10.1016/j.cpc.2013.10.016} {\bibfield  {journal} {\bibinfo  {journal}
  {Comput. Phys. Commun.}\ }\textbf {\bibinfo {volume} {185}},\ \bibinfo
  {pages} {960} (\bibinfo {year} {2014})},\ \Eprint
  {http://arxiv.org/abs/1305.0237} {arXiv:1305.0237 [hep-ph]} \BibitemShut
  {NoStop}%
\bibitem [{\citenamefont {Sjostrand}\ \emph {et~al.}(2015)\citenamefont
  {Sjostrand}, \citenamefont {Ask}, \citenamefont {Christiansen}, \citenamefont
  {Corke}, \citenamefont {Desai}, \citenamefont {Ilten}, \citenamefont
  {Mrenna}, \citenamefont {Prestel}, \citenamefont {Rasmussen},\ and\
  \citenamefont {Skands}}]{Sjostrand:2014zea}%
  \BibitemOpen
  \bibfield  {author} {\bibinfo {author} {\bibfnamefont {T.}~\bibnamefont
  {Sjostrand}}, \bibinfo {author} {\bibfnamefont {S.}~\bibnamefont {Ask}},
  \bibinfo {author} {\bibfnamefont {J.~R.}\ \bibnamefont {Christiansen}},
  \bibinfo {author} {\bibfnamefont {R.}~\bibnamefont {Corke}}, \bibinfo
  {author} {\bibfnamefont {N.}~\bibnamefont {Desai}}, \bibinfo {author}
  {\bibfnamefont {P.}~\bibnamefont {Ilten}}, \bibinfo {author} {\bibfnamefont
  {S.}~\bibnamefont {Mrenna}}, \bibinfo {author} {\bibfnamefont
  {S.}~\bibnamefont {Prestel}}, \bibinfo {author} {\bibfnamefont {C.~O.}\
  \bibnamefont {Rasmussen}}, \ and\ \bibinfo {author} {\bibfnamefont {P.~Z.}\
  \bibnamefont {Skands}},\ }\href {\doibase 10.1016/j.cpc.2015.01.024}
  {\bibfield  {journal} {\bibinfo  {journal} {Comput. Phys. Commun.}\ }\textbf
  {\bibinfo {volume} {191}},\ \bibinfo {pages} {159} (\bibinfo {year}
  {2015})},\ \Eprint {http://arxiv.org/abs/1410.3012} {arXiv:1410.3012
  [hep-ph]} \BibitemShut {NoStop}%
\bibitem [{\citenamefont {Buckley}\ \emph {et~al.}(2013)\citenamefont
  {Buckley}, \citenamefont {Butterworth}, \citenamefont {Lonnblad},
  \citenamefont {Grellscheid}, \citenamefont {Hoeth}, \citenamefont {Monk},
  \citenamefont {Schulz},\ and\ \citenamefont {Siegert}}]{Buckley:2010ar}%
  \BibitemOpen
  \bibfield  {author} {\bibinfo {author} {\bibfnamefont {A.}~\bibnamefont
  {Buckley}}, \bibinfo {author} {\bibfnamefont {J.}~\bibnamefont
  {Butterworth}}, \bibinfo {author} {\bibfnamefont {L.}~\bibnamefont
  {Lonnblad}}, \bibinfo {author} {\bibfnamefont {D.}~\bibnamefont
  {Grellscheid}}, \bibinfo {author} {\bibfnamefont {H.}~\bibnamefont {Hoeth}},
  \bibinfo {author} {\bibfnamefont {J.}~\bibnamefont {Monk}}, \bibinfo {author}
  {\bibfnamefont {H.}~\bibnamefont {Schulz}}, \ and\ \bibinfo {author}
  {\bibfnamefont {F.}~\bibnamefont {Siegert}},\ }\href {\doibase
  10.1016/j.cpc.2013.05.021} {\bibfield  {journal} {\bibinfo  {journal}
  {Comput. Phys. Commun.}\ }\textbf {\bibinfo {volume} {184}},\ \bibinfo
  {pages} {2803} (\bibinfo {year} {2013})},\ \Eprint
  {http://arxiv.org/abs/1003.0694} {arXiv:1003.0694 [hep-ph]} \BibitemShut
  {NoStop}%
\bibitem [{\citenamefont {Andersen}\ \emph {et~al.}(2013)\citenamefont
  {Andersen} \emph {et~al.}}]{Heinemeyer:2013tqa}%
  \BibitemOpen
  \bibfield  {author} {\bibinfo {author} {\bibfnamefont {J.~R.}\ \bibnamefont
  {Andersen}} \emph {et~al.} (\bibinfo {collaboration} {LHC Higgs Cross Section
  Working Group}),\ }\href {\doibase 10.5170/CERN-2013-004} {\  (\bibinfo
  {year} {2013}),\ 10.5170/CERN-2013-004},\ \Eprint
  {http://arxiv.org/abs/1307.1347} {arXiv:1307.1347 [hep-ph]} \BibitemShut
  {NoStop}%
\bibitem [{\citenamefont {Hamilton}\ \emph {et~al.}(2012)\citenamefont
  {Hamilton}, \citenamefont {Nason},\ and\ \citenamefont
  {Zanderighi}}]{Hamilton:2012np}%
  \BibitemOpen
  \bibfield  {author} {\bibinfo {author} {\bibfnamefont {K.}~\bibnamefont
  {Hamilton}}, \bibinfo {author} {\bibfnamefont {P.}~\bibnamefont {Nason}}, \
  and\ \bibinfo {author} {\bibfnamefont {G.}~\bibnamefont {Zanderighi}},\
  }\href {\doibase 10.1007/JHEP10(2012)155} {\bibfield  {journal} {\bibinfo
  {journal} {JHEP}\ }\textbf {\bibinfo {volume} {10}},\ \bibinfo {pages} {155}
  (\bibinfo {year} {2012})},\ \Eprint {http://arxiv.org/abs/1206.3572}
  {arXiv:1206.3572 [hep-ph]} \BibitemShut {NoStop}%
\bibitem [{\citenamefont {Boughezal}\ \emph
  {et~al.}(2015{\natexlab{a}})\citenamefont {Boughezal}, \citenamefont {Caola},
  \citenamefont {Melnikov}, \citenamefont {Petriello},\ and\ \citenamefont
  {Schulze}}]{Boughezal:2015dra}%
  \BibitemOpen
  \bibfield  {author} {\bibinfo {author} {\bibfnamefont {R.}~\bibnamefont
  {Boughezal}}, \bibinfo {author} {\bibfnamefont {F.}~\bibnamefont {Caola}},
  \bibinfo {author} {\bibfnamefont {K.}~\bibnamefont {Melnikov}}, \bibinfo
  {author} {\bibfnamefont {F.}~\bibnamefont {Petriello}}, \ and\ \bibinfo
  {author} {\bibfnamefont {M.}~\bibnamefont {Schulze}},\ }\href {\doibase
  10.1103/PhysRevLett.115.082003} {\bibfield  {journal} {\bibinfo  {journal}
  {Phys. Rev. Lett.}\ }\textbf {\bibinfo {volume} {115}},\ \bibinfo {pages}
  {082003} (\bibinfo {year} {2015}{\natexlab{a}})},\ \Eprint
  {http://arxiv.org/abs/1504.07922} {arXiv:1504.07922 [hep-ph]} \BibitemShut
  {NoStop}%
\bibitem [{\citenamefont {Boughezal}\ \emph
  {et~al.}(2015{\natexlab{b}})\citenamefont {Boughezal}, \citenamefont {Focke},
  \citenamefont {Giele}, \citenamefont {Liu},\ and\ \citenamefont
  {Petriello}}]{Boughezal:2015aha}%
  \BibitemOpen
  \bibfield  {author} {\bibinfo {author} {\bibfnamefont {R.}~\bibnamefont
  {Boughezal}}, \bibinfo {author} {\bibfnamefont {C.}~\bibnamefont {Focke}},
  \bibinfo {author} {\bibfnamefont {W.}~\bibnamefont {Giele}}, \bibinfo
  {author} {\bibfnamefont {X.}~\bibnamefont {Liu}}, \ and\ \bibinfo {author}
  {\bibfnamefont {F.}~\bibnamefont {Petriello}},\ }\href {\doibase
  10.1016/j.physletb.2015.06.055} {\bibfield  {journal} {\bibinfo  {journal}
  {Phys. Lett.}\ }\textbf {\bibinfo {volume} {B748}},\ \bibinfo {pages} {5}
  (\bibinfo {year} {2015}{\natexlab{b}})},\ \Eprint
  {http://arxiv.org/abs/1505.03893} {arXiv:1505.03893 [hep-ph]} \BibitemShut
  {NoStop}%
\bibitem [{\citenamefont {Anastasiou}\ \emph {et~al.}(2015)\citenamefont
  {Anastasiou}, \citenamefont {Duhr}, \citenamefont {Dulat}, \citenamefont
  {Herzog},\ and\ \citenamefont {Mistlberger}}]{Anastasiou:2015ema}%
  \BibitemOpen
  \bibfield  {author} {\bibinfo {author} {\bibfnamefont {C.}~\bibnamefont
  {Anastasiou}}, \bibinfo {author} {\bibfnamefont {C.}~\bibnamefont {Duhr}},
  \bibinfo {author} {\bibfnamefont {F.}~\bibnamefont {Dulat}}, \bibinfo
  {author} {\bibfnamefont {F.}~\bibnamefont {Herzog}}, \ and\ \bibinfo {author}
  {\bibfnamefont {B.}~\bibnamefont {Mistlberger}},\ }\href {\doibase
  10.1103/PhysRevLett.114.212001} {\bibfield  {journal} {\bibinfo  {journal}
  {Phys. Rev. Lett.}\ }\textbf {\bibinfo {volume} {114}},\ \bibinfo {pages}
  {212001} (\bibinfo {year} {2015})},\ \Eprint
  {http://arxiv.org/abs/1503.06056} {arXiv:1503.06056 [hep-ph]} \BibitemShut
  {NoStop}%
\bibitem [{\citenamefont {van Hameren}(2011)}]{vanHameren:2010cp}%
  \BibitemOpen
  \bibfield  {author} {\bibinfo {author} {\bibfnamefont {A.}~\bibnamefont {van
  Hameren}},\ }\href {\doibase 10.1016/j.cpc.2011.06.011} {\bibfield  {journal}
  {\bibinfo  {journal} {Comput. Phys. Commun.}\ }\textbf {\bibinfo {volume}
  {182}},\ \bibinfo {pages} {2427} (\bibinfo {year} {2011})},\ \Eprint
  {http://arxiv.org/abs/1007.4716} {arXiv:1007.4716 [hep-ph]} \BibitemShut
  {NoStop}%
\bibitem [{\citenamefont {Alwall}\ \emph {et~al.}(2014)\citenamefont {Alwall},
  \citenamefont {Frederix}, \citenamefont {Frixione}, \citenamefont {Hirschi},
  \citenamefont {Maltoni}, \citenamefont {Mattelaer}, \citenamefont {Shao},
  \citenamefont {Stelzer}, \citenamefont {Torrielli},\ and\ \citenamefont
  {Zaro}}]{Alwall:2014hca}%
  \BibitemOpen
  \bibfield  {author} {\bibinfo {author} {\bibfnamefont {J.}~\bibnamefont
  {Alwall}}, \bibinfo {author} {\bibfnamefont {R.}~\bibnamefont {Frederix}},
  \bibinfo {author} {\bibfnamefont {S.}~\bibnamefont {Frixione}}, \bibinfo
  {author} {\bibfnamefont {V.}~\bibnamefont {Hirschi}}, \bibinfo {author}
  {\bibfnamefont {F.}~\bibnamefont {Maltoni}}, \bibinfo {author} {\bibfnamefont
  {O.}~\bibnamefont {Mattelaer}}, \bibinfo {author} {\bibfnamefont {H.~S.}\
  \bibnamefont {Shao}}, \bibinfo {author} {\bibfnamefont {T.}~\bibnamefont
  {Stelzer}}, \bibinfo {author} {\bibfnamefont {P.}~\bibnamefont {Torrielli}},
  \ and\ \bibinfo {author} {\bibfnamefont {M.}~\bibnamefont {Zaro}},\ }\href
  {\doibase 10.1007/JHEP07(2014)079} {\bibfield  {journal} {\bibinfo  {journal}
  {JHEP}\ }\textbf {\bibinfo {volume} {07}},\ \bibinfo {pages} {079} (\bibinfo
  {year} {2014})},\ \Eprint {http://arxiv.org/abs/1405.0301} {arXiv:1405.0301
  [hep-ph]} \BibitemShut {NoStop}%
\bibitem [{\citenamefont {Farina}\ \emph {et~al.}(2013)\citenamefont {Farina},
  \citenamefont {Grojean}, \citenamefont {Maltoni}, \citenamefont {Salvioni},\
  and\ \citenamefont {Thamm}}]{Farina:2012xp}%
  \BibitemOpen
  \bibfield  {author} {\bibinfo {author} {\bibfnamefont {M.}~\bibnamefont
  {Farina}}, \bibinfo {author} {\bibfnamefont {C.}~\bibnamefont {Grojean}},
  \bibinfo {author} {\bibfnamefont {F.}~\bibnamefont {Maltoni}}, \bibinfo
  {author} {\bibfnamefont {E.}~\bibnamefont {Salvioni}}, \ and\ \bibinfo
  {author} {\bibfnamefont {A.}~\bibnamefont {Thamm}},\ }\href {\doibase
  10.1007/JHEP05(2013)022} {\bibfield  {journal} {\bibinfo  {journal} {JHEP}\
  }\textbf {\bibinfo {volume} {05}},\ \bibinfo {pages} {022} (\bibinfo {year}
  {2013})},\ \Eprint {http://arxiv.org/abs/1211.3736} {arXiv:1211.3736
  [hep-ph]} \BibitemShut {NoStop}%
\end{thebibliography}%

\end{document}